\documentstyle[multicol,aps]{revtex}

\def\beginwide{
        \end{multicols} \vspace*{-0.5cm} \noindent
        \rule{3.5in}{.1mm}\rule{.1mm}{5mm} \widetext \medskip }
\def\beginwidetop{
        \end{multicols} \vspace*{-0.5cm} \noindent
        \widetext \medskip }
\def\endwide{
        \hspace*{3.35in}~\rule[-5mm]{.1mm}{5mm}\rule{3.5in}{.1mm}
        \begin{multicols}{2} \vspace*{-1.0cm} \noindent }
\def\endwidebottom{
        \begin{multicols}{2} \vspace*{-1.0cm} \noindent }

\draft
\begin{document}

\title{
	Noisy Kuramoto-Sivashinsky equation for an erosion model}

\author{Kent B{\ae}kgaard Lauritsen,$^{(1,3)}$
	Rodolfo Cuerno,$^{(2,3)}$ and 
	Hern\'an A. Makse$^{(3)}$}

\address{
$^{(1)}$ Niels Bohr Institute, Center for Chaos and Turbulence Studies,
        Blegdamsvej 17, 2100 Copenhagen \O, Denmark
\\
$^{(2)}$ Departamento de Matem\'aticas, Universidad Carlos III de Madrid, 
	Butarque 15, 28911 Legan\'es, Spain
\\
$^{(3)}$ Center for Polymer Studies and Department of Physics, 
	Boston University, Boston, Massachusetts 02215
}

\date{\today}

\maketitle
\medskip

\begin{abstract}
We derive the continuum equation for a discrete model for ion sputtering.
We follow an approach based on the master equation,
and discuss how it can be truncated to a
Fokker-Planck equation and mapped to a discrete Langevin equation.
By taking the continuum limit, we arrive at the
Kuramoto-Sivashinsky equation with a stochastic noise term.

\end{abstract}

%

\begin{multicols}{2}

\section{Introduction}
\label{sec:intro}

The field of nonequilibrium interface growth is currently being
investigated by a variety of approaches
\cite{krug-spohn:1991,barabasi-stanley:1995,hh-zhang:1995}.
One direction of research is the formulation of 
Langevin equations which are studied by means of analytical and
numerical methods.
Typically, the terms appearing in Langevin equations
are derived by the use of symmetry arguments.
A different approach consists in the formulation of discrete models
which contain the relevant physical
mechanisms present in the problem at hand.
The models are studied by computer simulations in order to determine
which universality class they belong to.
One can try to relate the models to continuum descriptions. In many
occasions such an identification is indirect since it is 
difficult to derive the continuum description for a given set of
microscopic growth rules.

There exist cases in which it is possible to derive 
Langevin equations describing various discrete growth models
\cite{plischke-etal:1987,racz-etal:1991,vvedensky-etal:1993,park-kahng:1995}.
The approach followed in these cases is based on the master equation
which determines the evolution of the joint
probability density $P(H,t)$, where $H$ specifies the interface
configuration and $t$ denotes time.
The master equation reads
\begin{equation}
        \frac{\partial P(H,t)}{\partial t}
                = \sum_{H'} W(H'\!,H)P(H'\!,t) - \sum_{H'} W(H,H')P(H,t) ,
                                                \label{eq:me}
\end{equation}
where $W(H,H')$ denotes the transition rate per unit time from
configuration $H$ to $H'$, and the sums are over all configurations $H'$. 
One can study the associated Fokker-Planck equation and
therefrom derive a continuum Langevin equation for the surface height
provided the fluctuations in the system are not too large, and provided
the continuum limit can be justified \cite{continuum-remark}.
An important feature of the master equation approach is that
it provides a relation between the fluctuations (noise)
in the system and the parameters in the model.

In the present paper, we follow the master equation
approach in order to derive the continuum equation
for a simple pattern forming model recently introduced 
to study the morphology of surfaces eroded by ion sputtering
\cite{cuerno-etal:1995}.
The equation which we obtain for the evolution of the height profile
$h(x,t)$ is a noisy version of the
Kuramoto-Sivashinsky (KS) equation and reads
\begin{equation}
        \frac{\partial h}{\partial t}
             = v_0
                - |\nu | \nabla^2 h
                -\kappa (\nabla^2)^2 h
                + \frac{\lambda}{2} (\nabla h)^2
                + \eta(x,t) ,
                                        \label{eq:ks}
\end{equation}
where $v_0$, $\nu$, $\kappa$, and $\lambda$ are
constants, and $\eta(x,t)$ is a Gaussian white noise.
The KS equation (obtained when $\eta=0$)
\cite{ks:1977} is often considered a
paradigm of spatiotemporal chaos and appears frequently in studies of
pattern formation \cite{cross-hohenberg:1993}.
An initially flat
one-dimensional interface described by the KS equation
evolves in time from an almost periodic morphology at early times
to a rough surface at late times
\cite{zaleski:1989,sneppen-etal:1992,lvov-etal:1993}
described by the Kardar-Parisi-Zhang (KPZ) equation 
\cite{kpz:1986}.

The model for ion sputtering in Ref.\ \cite{cuerno-etal:1995}
shows an initial periodic morphology and a late-time KPZ scaling regime
which are similar to the behavior of the KS equation.
However, the KS equation is a deterministic equation whereas the
erosion model is inherently stochastic.
In \cite{cuerno-etal:1995} the early and late time dynamics of the
erosion model were numerically studied with the conclusion that they are
the same as those obtained from the noisy KS equation
\cite{karma-misbah:1993}.
Here, we confirm this result by showing analytically that the noisy KS
equation yields the continuum description of the erosion model.
By using the master equation approach, we determine the contributions
of the intrinsic noise to the evolution equation for this simple
pattern forming model.

\section{Erosion Model}
\label{sec:model}

The model introduced in Ref.\ \cite{cuerno-etal:1995} considers an
interface in $1+1$ dimensions described by the height
variable $h_i(t)$, $i=1,\ldots,L$, where $L$ is the system size.
The sites below the interface are occupied with particles,
whereas the sites above are empty.
The lattice unit distance along the horizontal direction is denoted by $a$.
The derivation of the noisy KS equation can be generalized to $d+1$ dimensions
but in the following we discuss the $1+1$ dimensional case.

The two basic physical mechanisms acting on the surface are
erosion and diffusion.
In the model, a particle at the interface is chosen randomly. 
Then,
the particle is subjected to an erosion rule with probability $f$,
and to a diffusion rule with probability $1-f$.

The erosion rule is as follows: the particle at $h_i$ is eroded
(the corresponding lattice position emptied) with probability
$P_e Y_i$ \cite{model}.
The quantity $P_e$ is computed as $1/7$ times the number of occupied sites
in a box of size $3\times 3$ centered at the chosen site,
i.e., $P_e$ is the number of occupied nearest and next-nearest neighbors of
the particle at $h_i$, and takes the values $1/7, 2/7, \ldots, 1$.
The definition of $P_e$ accounts for the
unstable erosion mechanism which exists in the physical systems due to
the finite penetration depth of the bombarding ions into the eroded substrate
\cite{sigmund}.
The efficiency of the
sputtering process is measured by the sputtering yield 
\cite{carter}
\begin{equation}
	Y_i = Y(\varphi_i) = y_0 + y_1 \, \varphi_{i}^{2} 
				+ y_2 \, \varphi_{i}^{4},
					\label{eq:Y}
\end{equation}
with the local slope 
$\varphi_{i} = \tan^{-1}((h_{i+1}-h_{i-1})/2a)$.
In \cite{cuerno-etal:1995} $y_0$, $y_1$, and $y_2$ were chosen such that
$Y_i(0)=0.5$,
$Y_i(\pi/2)=0$, and with a maximum value $Y_i(\varphi_m)=1$ for
$\varphi_m = 1$ rad.
In general one would merely require that $y_0>0$, $y_1>0$, and $y_2<0$
\cite{carter}.

The surface diffusion rule moves a particle at the interface from
the top of column $k$ to the top of a randomly chosen nearest-neighbor
column with the rates
\begin{equation}
	w_{k}^{\pm} 
		= \frac{1}{1+\exp(\beta \Delta {\cal H}_{k \to k\pm 1})},
					\label{eq:surf-diff}
\end{equation}
where the energy is 
$
	{\cal H} = ({J}/{b^2}) \sum_{i=1}^{L} (h_i - h_{i+1})^2 
$
\cite{siegert-plischke:1994},
and $\Delta {\cal H}_{k \to k\pm 1}$ is the energy difference between the
final and initial configurations; $J$ is a coupling constant,
$b$ is the unit lattice spacing in the height direction, and
$\beta$ is the inverse temperature. 

\section{Stochastic Formalism}
\label{sec:stochastic}

The interface configuration at a given time $t$ is
specified by the set of column variables $H=\{h_1, h_2, \ldots, h_L\}$.
The moments of the transition rates $W(H,H')$ in Eq.~(\ref{eq:me})
are defined as follows:
\begin{mathletters}
\label{eq:K-def}
\begin{eqnarray}
	K^{(1)}_{i} &=& \sum_{H'} ({h_i}'-h_i) W(H',H),
						\label{eq:K1-def}
							\\
	K^{(2)}_{i,j} &=& \sum_{H'} ({h_i}'-h_i)({h_j}'-h_j) W(H',H),
						\label{eq:K2-def}
							\\
	K^{(n)}_{i_1,\ldots,i_n} 
		&=& \sum_{H'} \left(\prod_{k=1}^{n}({h_k}'-h_k)\right) W(H',H).
						\label{eq:Kn-def}
\end{eqnarray}
\end{mathletters}
\noindent
The first step in order to derive the Langevin equation
is to transform the master equation (\ref{eq:me}) into a Kramers-Moyal
partial differential equation (see, e.g., \cite{risken:1989})
\begin{equation}
	\frac{\partial P}{\partial t}
		= 
		  \sum_{n=1}^{\infty} \frac{(-1)^n}{n!} 
		\frac{\partial^n}{\partial h_{i_1} \ldots \partial h_{i_n}}
			\left( K^{(n)}_{i_1,\ldots,i_n}P \right) ,
					\label{eq:KM}
\end{equation}
where a sum over repeated indices is assumed.

Next, one identifies a parameter $\Omega$ such that 
$\Omega\to\infty$ denotes the macroscopic limit.
%
%
For $|h_i'-h_i|\sim O(1/\Omega)$, the ratio
$K^{(n+1)}/K^{(n)}$ will typically be of order $O(1/\Omega)$ \cite{fox:1978}.
If the intrinsic fluctuations are sufficiently small 
then only the $K^{(1)}$ term in Eq.~(\ref{eq:KM})
will survive in the limit $\Omega \to \infty$
\cite{fox-keizer:1991}.
However, in our case the intrinsic fluctuations are relevant.
Then, in the limit $\Omega \to \infty$
the Kramers-Moyal expansion reduces to the Fokker-Planck equation
\begin{equation}
	\frac{\partial P}{\partial t}
		= -\frac{\partial}{\partial h_i}
			\left( K^{(1)}_{i}P \right)
		  +\frac{1}{2} \frac{\partial^2}{\partial h_i \partial h_j}
			\left( K^{(2)}_{i,j}P \right)  ,
					\label{eq:FP}
\end{equation}
which is obtained from Eq.~(\ref{eq:KM}) by keeping only
the $K^{(1)}$ and $K^{(2)}$ terms \cite{FP-correction}.


The Fokker-Planck equation can be used in the further
analysis of the system. However, only average quantities such as, e.g.,
$
        \left< h_i(t) \right> = \sum_{H} h_i P(H,t)  ,
$
can be calculated. Therefore it is generally more convenient to recast  
Eq.~(\ref{eq:FP}) in
an equivalent Langevin form for the height $h_i(t)$.
In the Stratonovich interpretation, the Langevin equation associated to
the Fokker-Planck equation is 
\begin{equation}
        \frac{\partial h_i}{\partial t} = K^{(1)}_{i} + \eta_i,
					\label{eq:langevin}
\end{equation}
where terms of order $O(1/\Omega)$ have been neglected, cf.\ 
\cite{fox-keizer:1991}. In Eq.~(\protect\ref{eq:langevin}), 
$h_i$ is a continuous variable describing the dynamics of the fluctuations
of the height configurations $\{h_1(t),\ldots,h_L(t)\}$, as obtained in the
macroscopic limit $\Omega \rightarrow \infty$, cf.\ \cite{fox:1978} 
(see also \cite{racz-etal:1991,siegert-plischke:1994}).
The term $\eta_i$ is a Gaussian white noise with average value 
equal to zero, and variance
\begin{equation}
	\left< \eta_i(t)\eta_j(t')\right> = K^{(2)}_{i,j} \delta(t-t').
					\label{eq:noise}
\end{equation}

\section{Discrete Langevin Equation}
\label{sec:discrete}

We now apply the formalism described in the previous
section to the erosion model in \cite{cuerno-etal:1995}.
The transition rate for the erosion rule reads
\begin{equation}
	W_e(H,H') =  \frac{f}{\tau} \sum_{k} P_e Y_k 
			 \delta(h_{k}',h_k-b) 
			\prod_{j\neq k} \delta(h_{j}',h_j)  ,
                                        \label{eq:w-erosion}
\end{equation}
where $\tau$ is the time scale.
The erosion probability $P_e$ can be expressed in different forms.
We choose
\begin{equation}
	P_e = \frac{1}{7} \left( 5 + \frac{a_0}{a^2} \nabla^2 h_i +
			{a_1} \, \Theta \right) ,
							\label{eq:P_e}
\end{equation}
with the discrete Laplacian 
\mbox{$
	\nabla^2 h_i = h_{i+1} - 2h_i + h_{i-1}    
$},
and constants $a_0$ and $a_1$.
In Eq.~(\ref{eq:P_e}) the Laplacian accounts for the physical mechanism of
the box rule, and 
the $\Theta$ term accounts for the finite size of the box.
Defining 
	$\theta(x)=1$ for $x \ge 0$, 
and 
	$\theta(x)=0$ for $x < 0$,
the $\Theta$ term can be written
\begin{eqnarray}
	\Theta = &-&
		\theta\left({h_{i-1}-h_i-2b}\right)
			\left[{h_{i-1}-h_i-b}\right]
							  \nonumber \\
	    	&-&
		\theta\left({h_{i+1}-h_i-2b}\right)
			\left[{h_{i+1}-h_i-b}\right] 
							  \nonumber \\
	    	&+&
		\theta\left({h_i-h_{i-1}-3b}\right)
			\left[{h_i-h_{i-1}-2b}\right]
							  \nonumber \\
	    	&+&
		\theta\left({h_i-h_{i+1}-3b}\right)
			\left[{h_i-h_{i+1}-2b}\right]
		.
							\label{eq:Theta}
\end{eqnarray}
The $\Theta$ term in Eqs.~(\ref{eq:P_e}) and (\ref{eq:Theta})
becomes effectively zero in the hydrodynamic limit of the model where
the surface roughens in a way consistent with the KPZ universality class
and the slopes are small along the interface.
Also, from the numerical simulation \cite{cuerno-etal:1995}
it is known that the qualitative behavior of the model does not change
if the size of the box is enlarged \cite{box-rule}.
The sputtering yield is obtained from Eq.~(\ref{eq:Y}) as 
\begin{equation}
	Y_i 
	= y_0 + \frac{y_1}{a^2} (\nabla h_i)^2 
			+ \frac{y_2-2y_1/3}{a^4} (\nabla h_i)^4 + \ldots
		,
					\label{eq:Y2}
\end{equation}
with the discrete gradient
\mbox{$
	\nabla h_i = (h_{i+1}-h_{i-1})/{2}   
$}.

The transition rate for the diffusion rule reads
\beginwide
\begin{eqnarray}
	W_d(H,H') &=& 
		\frac{1-f}{2\tau} \sum_{k}
		\left[
		w_{k}^{+} \delta(h_{k}',h_k-b)\delta(h_{k+1}',h_{k+1}+b)
		+
		w_{k+1}^{-} \delta(h_{k}',h_k+b)\delta(h_{k+1}',h_{k+1}-b)
		\right]
		\prod_{j\neq k,k+1} \delta(h_{j}',h_j) 
		 ,
                                        \label{eq:w-diff}
\end{eqnarray}
where $w_{k}^{\pm}$ are defined in Eq.~(\ref{eq:surf-diff}).
By expanding the transition rates $w_{k}^{\pm}$, we find 
\begin{equation}
	w_{k}^{\pm} = \frac{1}{1+q} 
		-\frac{2\beta Jq/b}{(1+q)^2}
			\left( \nabla^2 h_k - \nabla^2 h_{k\pm 1} \right)
		+\sum_{n=2} 
			\frac{C_n(q)}{b^n} 
			\left( \nabla^2 h_k - \nabla^2 h_{k\pm 1}\right)^{n} ,
                                        \label{eq:w-diff2}
\end{equation}

\endwide
where $q \equiv \exp(6\beta J)$, and $C_n(q)$ are numerical constants
whose exact value will not be needed in what follows.

Using that $W(H,H')=W_e(H,H')+W_d(H,H')$,
and expression (\ref{eq:K-def}), 
we obtain to lowest order the following values for the transition moments:
\begin{mathletters}
                     \label{eq:K}
\begin{eqnarray}
	K_{i}^{(1)} &=& - \frac{fb}{\tau} P_e Y_i
		-\frac{(1-f)\beta Jq}{\tau(1+q)^2} \nabla^2 (\nabla^2 h_i) ,
                                                        \label{eq:K1}  \\
	&&\nonumber\\
	K_{i,j}^{(2)} &=& -\frac{(1-f)b^2}{\tau(1+q)} 
				\nabla^2 \delta_{ij} 
			+\frac{fb^2}{\tau} 
			P_e Y_i \delta_{ij}
			.
                                        \label{eq:K2}
\end{eqnarray}
\end{mathletters}
\noindent
Here, 
$\nabla^2 \delta_{ij} = \delta_{i+1,j} - 2\delta_{i,j} + \delta_{i-1,j}$.
Higher order terms of the form
\mbox{$(\nabla^2 h_{i} - \nabla^2 h_{i \pm 1})^{2n+1}$} have been omitted
(in section \ref{sec:cont}
we will argue that such terms are irrelevant in the renormalization
group (RG) sense for the scaling properties of the interface).
The quantity $P_e Y_i$ reads
\begin{equation}
	P_e Y_i = \frac{5y_0}{7}
                        +\frac{y_0 a_0}{7a^2} \nabla^2 h_i
                        +\frac{5y_1}{7a^2} (\nabla h_i)^2
			,
\end{equation}
with additional terms of the form 
$(\nabla h_i)^{2m} \nabla^2 h_i $.

The different terms appearing in Eq.~(\ref{eq:K})
give rise to the following contributions in the Langevin equation
(\ref{eq:langevin}):
an additive constant which corresponds to the average velocity of erosion
for a flat interface,
a negative Laplacian which reflects the unstable nature of the erosion rule,
and a $(\nabla h_i)^2$ nonlinearity of the KPZ type 
describing lateral motion of the interface \cite{kpz:1986};
the nonlinearity originates from the coefficient $y_1$
in expression (\ref{eq:Y}) for the sputtering yield. 
In addition, the diffusion rule contributes to Eq.~(\ref{eq:langevin})
with the linear term $\nabla^4 h_i$
(plus higher order nonlinearities).
As demonstrated in Ref.\ \cite{siegert-plischke:1994} by numerical
simulations and shown using a coarse-graining procedure,
the $\nabla^4 h_i$ term originates from the conserved nature
of the surface diffusion rule (\ref{eq:surf-diff})
\cite{racz-etal:1991,siegert-plischke:1994}.

\section{Continuous Langevin Equation}
\label{sec:cont}

To obtain a continuous Langevin equation, we assume that the discrete
function $h_i(t)$ can be replaced by a smooth function $h(x,t)$
such that $h_i(t) = h(x=ia,t)$, and further that  
\begin{equation}
        h_{i\pm 1}(t) - h_i(t) = \sum_{n=1}^{\infty} \frac{(\pm a)^n}{n!}
		\left. \frac{\partial^n h}{\partial x^n} \right|_{x=ia} .
                                        \label{eq:smooth}
\end{equation}
A more rigorous way to arrive at the function $h(x,t)$ would be through
some coarse graining procedure which preserves the symmetry of the problem. 
However, as discussed in \cite{racz-etal:1991,siegert-plischke:1994},
this is a highly nontrivial task. 
Here we are concerned with the form of the relevant
terms in the equation of motion for $h(x,t)$,
which are expected to have the 
same form as those obtained by the use of (\ref{eq:smooth}). We will
determine them as the leading terms in an expansion in which we take the
parameters $a$ and $b$ to be small but nonzero \cite{a-b}.
Combining (\ref{eq:smooth})
with Eqs.\ (\ref{eq:langevin}), (\ref{eq:noise}) 
we obtain the continuous Langevin equation for $h(x,t)$:
\begin{mathletters}
                                                \label{eq:ks2c}
\begin{eqnarray}
        \frac{\partial h}{\partial t} 
             &=& v_0
		- |\nu | \nabla^2 h 
		-\kappa (\nabla^2)^2 h  
		+ \frac{\lambda}{2} (\nabla h)^2 
		+ \eta(x,t)  
					\label{eq:ks2ac}    \\
	&&\nonumber\\
	~~&&+ 
		c_1 (\nabla h)^{2} \nabla^2 h
		+ c_2 (\nabla^3 h)^{2} \nabla^4 h
		+ \ldots
		,
                                                \label{eq:ks2bc}
\end{eqnarray}
\end{mathletters}
\noindent
where now $\nabla \equiv \partial/\partial x$.
The noise $\eta(x,t)$ has correlations 
\begin{eqnarray}
        \langle \eta(x,t) \eta(x',t') \rangle 
	    &=&  
		\left[ 2D - 2D_{d} \nabla^2 
		\right]
		\delta(x-x')\delta(t-t') 
						\nonumber \\
	    &+& \! 
		\left[ D_1 \nabla^2 h + D_2 (\nabla h)^2 
		\right]
		\delta(x-x')\delta(t-t') 
		. \!\!\!
                                        \label{eq:ks-noisec}
\end{eqnarray}

The coefficients in Eqs.~(\ref{eq:ks2c})-(\ref{eq:ks-noisec}) read
\begin{eqnarray}
	v_0 &=& -\frac{5f y_0 b}{7\tau} ,
	~~~~
	|\nu| = \frac{f y_0 a_0 b}{7\tau} ,   
	~~~~
        \kappa = \frac{(1-f)\beta Jqa^4}{\tau(1+q)^2} ,
	 					\nonumber \\
	\lambda &=& -\frac{10 f y_1 b}{7\tau} , 
	~~~~
	D     = \frac{5f y_0 a b^2}{14\tau} , 
	~~~~
	D_{d} = \frac{(1-f)a^3 b^2}{2 \tau (1+q)^2} ,
	 					\nonumber \\
	D_1  &=& \frac{f y_0 a a_0 b^2}{7\tau} , 
	~~~~
	D_2   = \frac{5f y_1 a b^2}{7\tau} . 
						\label{eq:parameters} 
\end{eqnarray}
The terms in Eq.~(\ref{eq:ks2ac})
constitute the noisy Kuramoto-Sivashinsky equation (\ref{eq:ks}), where one
notes the presence of a negative Laplacian and a KPZ nonlinearity.
The presence of such a nonlinearity is known to determine the scaling
behavior of the noisy KS equation as shown by an RG analysis
\cite{cuerno-lauritsen:1995}.
Using the values of the roughness and dynamic exponents at the KPZ fixed
point \cite{kpz:1986},
the terms appearing in Eq.~(\ref{eq:ks2bc})
can be shown to be irrelevant.
A similar RG argument shows the irrelevance of the conserved
term $D_{d}$ in the noise correlator (\ref{eq:ks-noisec}) 
in the presence of the
shot noise $D$, as well as the irrelevance of the multiplicative
contributions $D_1$ and $D_2$.


\section{Conclusions}
\label{sec:conclusions}

In summary, we have derived the continuum equation for a
microscopic model for ion sputtering.
The resulting equation is a noisy version
of the Kuramoto-Sivashinsky equation. Such an equation exhibits
the same scaling properties as the Kardar-Parisi-Zhang equation
and this result is in nice agreement with the simulations of the erosion
model performed in Ref.\ \cite{cuerno-etal:1995}.
The master equation approach has allowed us to derive the
form of the stochastic noise in the Langevin equation
for a simple pattern forming system.

\section*{Acknowledgements}

We acknowledge discussions with S. Tomassone and H. E. Stanley.
K.~B.~L.
acknowledges the support from the Danish Natural Science Research Council,
and R.~C.
acknowledges support from Ministerio de Educaci\'on y Ciencia, Spain.
The Center for Polymer Studies is supported by NSF.

\end{multicols}




\begin{multicols}{2}

\end{multicols}


\begin{thebibliography}{99}

\vspace*{-1.5cm}

\bibitem{krug-spohn:1991}
        J. Krug and H. Spohn,
        ``Kinetic Roughening of Growing Surfaces'',
        in
        {\em Solids far from Equilibrium: Growth, Morphology and Defects},
        ed.\ C. Godr\`eche
        (Cambridge University Press, Cambridge, 1991).
\bibitem{barabasi-stanley:1995}
        A.-L. Barab\'asi and H. E. Stanley,
        {\em Fractal Concepts in Surface Growth}
        (Cambridge University Press, Cambridge, 1995).
\bibitem{hh-zhang:1995}
        T. Halpin-Healey and Y.-C. Zhang,
	{Phys. Rep.} {\bf 254}, 215 (1995).


\bibitem{plischke-etal:1987}
	M. Plischke, Z. R\'acz, D. Liu,
	Phys. Rev. B {\bf 35}, 3485 (1987).
\bibitem{racz-etal:1991}
	Z. R\'acz, M. Siegert, D. Liu, and  M. Plischke,
	Phys. Rev. A {\bf 43}, 5275 (1991).
\bibitem{vvedensky-etal:1993}
	D. D. Vvedensky, A. Zangwill, C. N. Luse, and M. R. Wilby,
	Phys. Rev. E {\bf 48}, 852 (1993).
\bibitem{park-kahng:1995}
	K. Park and B. Kahng,
	Phys. Rev. E {\bf 51}, 796 (1995).

\bibitem{continuum-remark}
	See Refs.\ \protect\cite{plischke-etal:1987,racz-etal:1991}
	wherein difficulties with respect to the continuum limit
	are discussed.

\bibitem{cuerno-etal:1995}
	R. Cuerno, H. A. Makse, S. Tomassone, S. T. Harrington, 
	and H. E. Stanley,
	Phys. Rev. Lett. {\bf 75}, 4464 (1995).

\bibitem{ks:1977}
	Y. Kuramoto and T. Tsuzuki, 
	{Prog. Theor. Phys.} {\bf 55}, 356 (1977);
	G. I. Sivashinsky, {Acta Astronaut.} {\bf 6}, 569 (1979).
\bibitem{cross-hohenberg:1993}
	M. C. Cross and P. C. Hohenberg,
	{Rev. Mod. Phys.} {\bf 65}, 851 (1993).

\bibitem{zaleski:1989}
	S. Zaleski, {Physica D} {\bf 34}, 427 (1989).
\bibitem{sneppen-etal:1992}
	K. Sneppen, J. Krug, M. H. Jensen, C. Jayaprakash, and T. Bohr,
	{Phys. Rev. A} {\bf 46}, R7351 (1992).
\bibitem{lvov-etal:1993}
	V. L'vov, V. V. Lebedev, M. Paton, and I. Procaccia, 
	{Nonlinearity} {\bf 6}, 25 (1993).

\bibitem{kpz:1986}
        M. Kardar, G. Parisi, and Y.-C. Zhang, 
	{Phys. Rev. Lett.} {\bf 56}, 889 (1986).

\bibitem{karma-misbah:1993}
	A. Karma and C. Misbah,
	{Phys. Rev. Lett.} {\bf 71}, 3810 (1993).

\bibitem{model}
	In Ref.\ \protect\cite{cuerno-etal:1995} the erosion rule
	was implemented in a different way which, however, 
	is equivalent to the form given here.
\bibitem{sigmund}
	P. Sigmund, 	
	J. Mat. Sci. {\bf 8}, 1545 (1973);
	R. M. Bradley and J. M. E. Harper, 
	{J. Vac. Sci. Technol. A} {\bf 6}, 2390 (1988);
	R. Cuerno and A.-L. Barab\'asi, 
	{Phys. Rev. Lett.} {\bf 74}, 4746 (1995).
\bibitem{carter}
	G. Carter, B. Navin\v{s}ek, and J. L. Whitton, in
	{\em Sputtering by Particle Bombardment}, R. Behrisch, ed.,
	Vol. II (Springer-Verlag, Heidelberg, 1983).
\bibitem{siegert-plischke:1994}
	M. Siegert and M. Plischke,
	Phys. Rev. E {\bf 50}, 917 (1994).


\bibitem{risken:1989}
        H. Risken,
        {\em The Fokker-Planck Equation: Methods of Solution and Applications},
        2nd ed.\ (Springer-Verlag, Heidelberg, 1989).
\bibitem{fox:1978}
	R. F. Fox,
	Phys. Rep. {\bf 48}, 179 (1978).
\bibitem{fox-keizer:1991}
	R. F. Fox and J. Keizer,
	Phys. Rev. A {\bf 43}, 1709 (1991).

\bibitem{FP-correction}
	The Fokker-Planck equation (\protect\ref{eq:FP}) will be 
	correct to order
	$O(\ln\Omega/\Omega)$, cf.\ \cite{fox-keizer:1991}.

\bibitem{box-rule} 
	In such a case, say for a $3 \times N$ box,
	the change in $P_e$ in Eq.~(\protect\ref{eq:P_e})
	affects merely the constant term and the $\Theta$ term 
	which becomes effectively zero for even earlier stages of the dynamics
	than in the $3 \times 3$ case.

\bibitem{cuerno-lauritsen:1995}
	R. Cuerno and K. B. Lauritsen,
	Phys. Rev. E {\bf 52}, 4853 (1995).

\bibitem{a-b}
	In the discrete model, $b$ is the analog of
	the finite penetration depth of the arriving ions into the solid
	substrate, while $a$ would correspond 
	to the width of the spatial distribution of energy deposited
	by a bombarding ion within distance $b$ from the surface.
	Both quantities represent a lower wavelength cutoff or spatial
	resolution whose typical values for physical systems are of the order 
	of $100$ \AA \protect\cite{sigmund}. 

\end{thebibliography}
\end{document}